\documentclass[a4paper,11pt]{article}
\pdfoutput=1 

\usepackage{jinstpub} 

\title{\boldmath Hyper-Kamiokande}

\author[a,b,c]{Yury Kudenko} 
 
\affiliation[a]{Institute for Nuclear Research of RAS,\\60 October Revolution Pr 7a,
Moscow, Russia}
\affiliation[b]{Moscow Institute of Physics and Technology, 
\\Institutskiy per. 9, Dolgoprudny, Moscow Region, Russia}
\affiliation[c]{Moscow  Engineering Physics Institute,\\Kashirskoe shosse 31, Moscow,  Russia}

\emailAdd{kudenko@inr.ru}

\abstract{A next generation water Cherenkov detector Hyper-Kamiokande  to be built in Japan is described. The main goals of this project include a sensitive measurement of  CP violation  in neutrino oscillations, a search for proton decay and study of solar, atmospherics and astrophysical neutrinos. Key features of the Hyper-Kamiokande detector are described. The main emphasis is put on large photosensors. The recent progress in development of near neutrino detectors is also presented. }


\collaboration[c]{on behalf of the Hyper-Kamiokande Proto-Collaboration}

\proceeding{International Conference "Instrumentation for Colliding Beam Physics" \\
 (INSTR2020) \\
  24 - 28  February, 2020 \\
  Novosibirsk, Russia }

\begin{document}
\maketitle
\flushbottom

\section{Introduction}
\label{sec:intro}
\begin{figure}
\centering\includegraphics[width=11cm,angle=0]{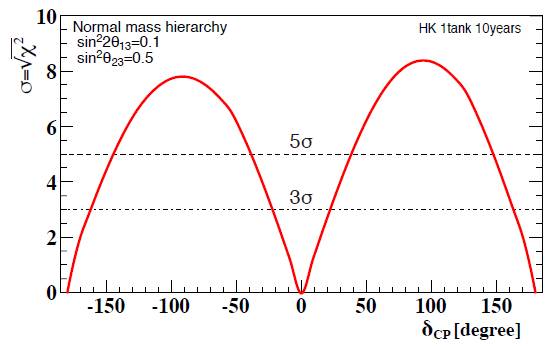}
\caption{Expected significance to exclude CP conservation ($\delta_{CP} = 0$ or $\pi$ ) for the  normal mass order. The significance is calculated as $\sqrt{\Delta\chi^2}$, where $\Delta\chi^2$  the   difference of $\chi^2$  for the trial value $\delta_{CP}$ and for $\delta_{CP} = 0 $ or $\pi$. The smaller value of difference is taken.}
\label{fig:hk_sensitivity}
\end{figure}

A next generation underground water Cherenkov detector Hyper-Kamiokande 
is being developed by an international collaboration as a leading 
worldwide experiment to  address fundamental  unsolved questions in particle
physics and cosmology~\cite{Abe:2018uyc,Abe:2015zbg}. It will be used as a far neutrino detector in the long baseline experiment with the  intensive neutrino and antineutrino beams from the upgraded Japan Proton Accelerator Research Complex (J-PARC). The main goal of these measurements is a sensitive search for CP violation in the leptonic sector of the Standard Model. As seen from Fig.~\ref{fig:hk_sensitivity},       
 which shows the expected significance to exclude $\delta_{CP} = 0$  or $\pi$ (the CP conserving cases) after 10 years of data taking,  CP violation  in neutrino oscillations can be observed with $\geq 5(3)\sigma$ significance for 57(80)\% of the possible values of $\delta_{CP}$. Exclusion of $\delta_{CP} = 0$ can be obtained with a significance of  $8\sigma$  in the case of maximal CP violation with $\delta_{CP} = -\pi/2$.  Hyper-Kamiokande  will also increase existing sensitivity to proton decay predicted in Grand Unified Theories by an order of magnitude. Atmospheric neutrinos will be used to study the neutrino mass hierarchy. Hyper-Kamiokande will provide the conclusive evidence of the day-night solar flux asymmetry and will make a  sensitive measurement of the solar neutrino spectrum upturn.   The excellent ability of Hyper-Kamiokande to distinguish the charge current $\nu_{\mu}$ and $\nu_e$ interactions allows the detector  to test the mass hierarchy in both  $\nu_{\mu} \to \nu_{\mu}$ and $\nu_{\mu} \to \nu_e$ channels.
In the case of a nearby Supernova, Hyper-Kamiokande will observe a large number of neutrino events, providing important experimental results  to understand the mechanism of the explosion. Hyper-Kamiokande will help to improve our understanding of some   phenomena in the Universe  by   detecting  astrophysical neutrinos from sources such as dark matter annihilation, gamma ray burst jets, and pulsar winds.
 
\section{Hyper-Kamiokande Detector}
\label{sec:detector}
\begin{figure}[htbp]
\centering 
\includegraphics[width=10cm]{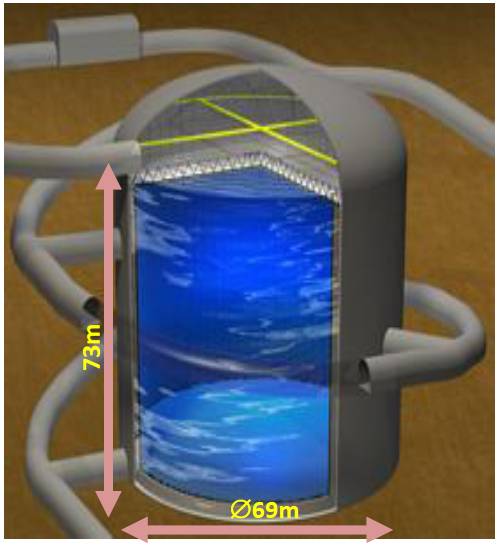} 
\caption{\label{fig:hyperk} Schematic view of the Hyper-Kamiokande detector.}
\end{figure}

Hyper-Kamiokande is based on the proven technology of the  highly successfull Super-Kamiokande detector~\cite{Fukuda:2002uc}.  The Hyper-Kamiokande detector (Fig.~\ref{fig:hyperk}) 
will be located in a  new cylindrical shaped cavern that  will be excavated at the Tochibora mine, about 8 km south of Super-Kamiokande,    with  an overburden of 1750 m.w.e.   It will consist of the  cylindrical tank (73 m high and 69 m in diameter) and have a total (fiducial) mass of 237 (187) kton, making it 5 (7.5) times larger than its predecessor Super-Kamiokande.   Similar to Super-Kamiokande, an outer detector with the layer width  of 1 m will help to constrain the external background. Hyper-Kamiokande will be the largest  underground water Cherenkov detector in the world and  will be instrumented with   40000 newly developed high-efficiency and high-resolution PMTs in the baseline design. The detector will be  filled with highly transparent ultra-pure water with  a light attenuation length  of above 100 meters is expected to  be achieved. 

\begin{figure}[htbp]
\centering 
\includegraphics[width=8cm]{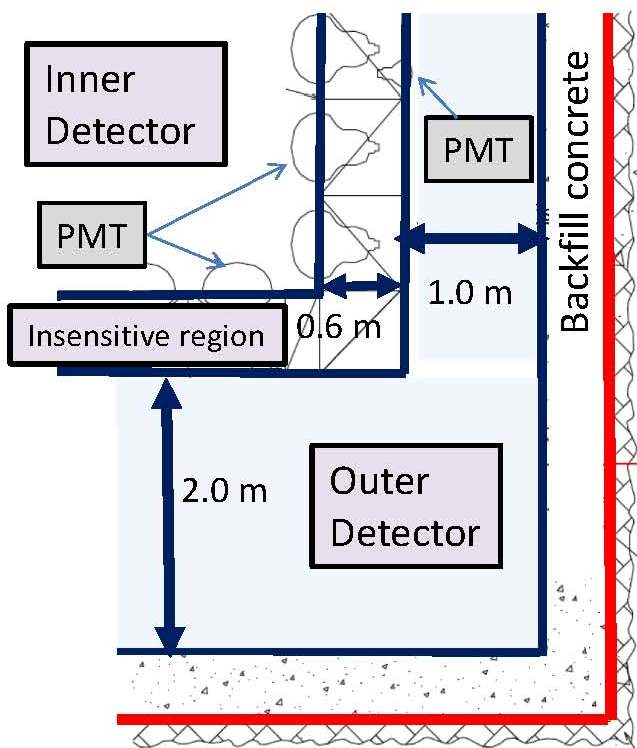} 
\caption{\label{fig:id_od} The Hyper-Kamiokande bottom region: the Inner Detector (ID) with the barrel and bottom regions of the Outer Detector 
(OD) are shown. }
\end{figure}

The water volume of the  tank contains two photo-sensitive segments optically separated  by a 60 cm thick insensitive region.  The inner segment called the Inner Detector (ID) has a cylindrical shape of 67 m in diameter and 69 m in height. This main active volume  is viewed by an array of inward-facing  40000 50 cm PMTs. The outer segment monitored by 
outward-facing  photosensors (15000 7.5 cm PMTs) called the Outer Detector (OD), which acts mainly as a veto  for entering particles such as cosmic  muons. The OD water thickness is 1 meter in the barrel region and 2 meters in the top and bottom regions as shown in Fig.~\ref{fig:id_od}.
Because  of the   lower density of OD photosensors, the photocathode coverage of the OD wall is expected to be   about 1\%. The photosensors for  ID and OD are mounted on a stainless steel supporting framework. The space between the ID PMTs is lined with opaque black sheets to prevent light leaks, while the gaps between the OD photosensors are lined with reflective sheets to enhance light collection in  OD. The design of the Hyper-Kamiokande water purification system will be based on the current Super-Kamiokande water system. In order to keep the water transparency the circulation of  water  should occur at a speed of  about 300 tons/hour. The radon concentration in the tank is expected to be  below 1mBq/m$^3$.
 
\section{New Photodetectors}
\label{sec:pmts}
\begin{figure}[htbp]
\centering 
\includegraphics[width=12cm]{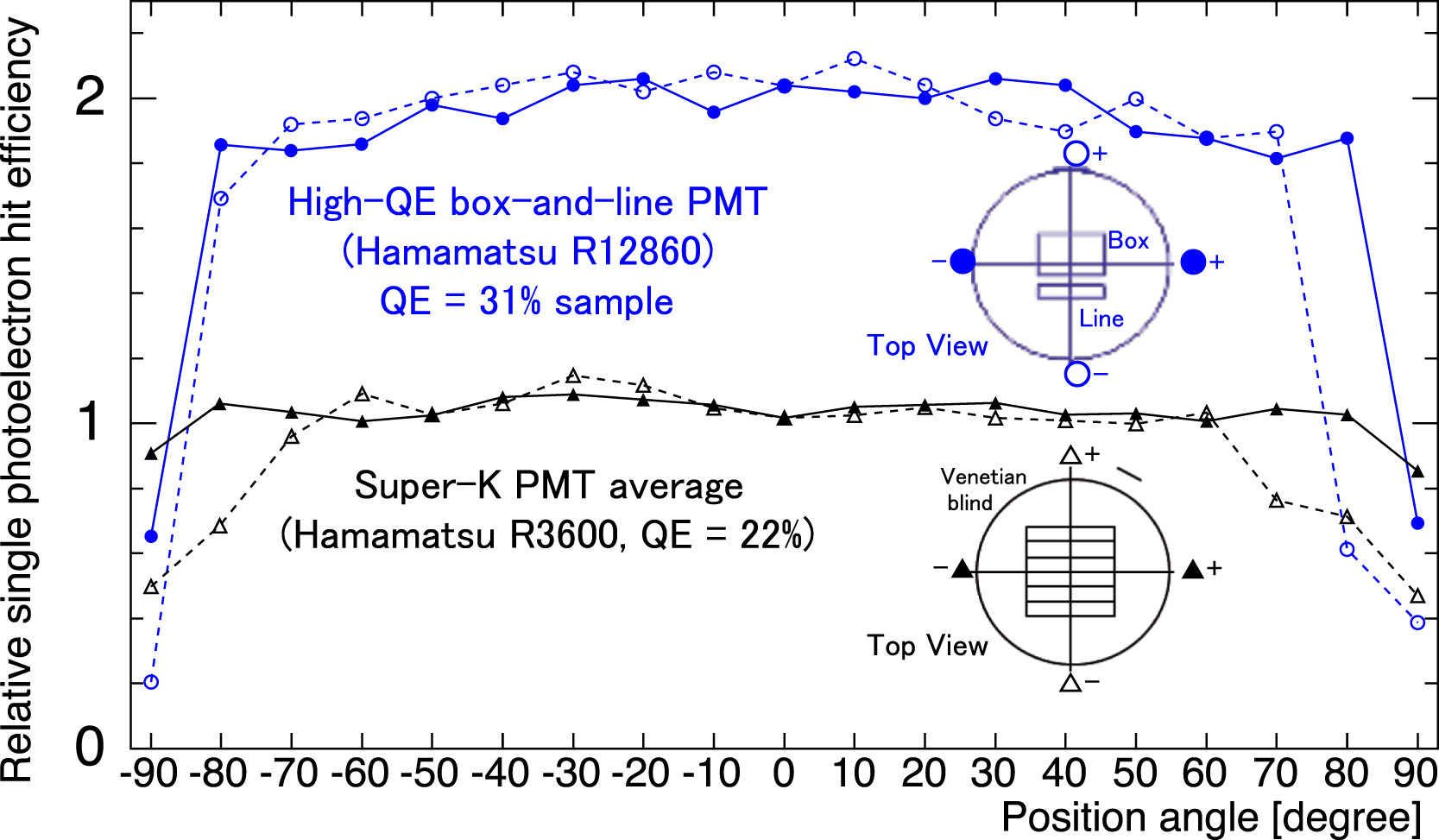} 
\caption{\label{fig:hyperk_pmt}The single photon detection efficiency as  a function of incident positions on the photocathode of the  Hamamatsu PMT R12860  developed  for Hyper-Kamiokande~\cite{Nishimura:2020eyq}. 
Also shown parameters of  the Super-Kamiokande  Hamamatsu  PMT R3600.  }
\end{figure}

In order to achieve  broad scientific goals, particles with a wide range of energies should be  reconstructed. The number of Cherenkov  photons that hit each photosensor ranges from one to several hundred. Thus, the photosensors are required to have a high photon detection efficiency, a  wide dynamic range, a low dark rate,  and a good linearity. The location of the neutrino interaction vertex is reconstructed using the Cherenkov photon arrival timing information at each PMT. Therefore, good timing resolution of  photosensors is essential, and the jitter of the transit time is required to be less than 3 ns ($1\sigma$) for a single photon. The dark rate of about 4 kHz at the  temperature of the Hyper-Kamiokande water is required for the solar neutrino measurements  with low energy threshold. To meet these requirements,   a  new 50 cm diameter Hamamatsu PMT R12860   with a Box{\&}Line dynode has been developed for Hyper-Kamiokande. The Box{\&}Line PMT~\cite{Nishimura:2020eyq} realized a high collection efficiency of photoelectrons  at the first box-shape dynode, and narrow timing variation by line-focused dynodes. It has a faster time response and a better collection efficiency compared to the the Venetian blind dynode Hamamatsu PMT  R3600 that has been used successfully in the Super-Kamiokande detector. It is expected that the required  dark rate of about 4 kHz will be obtained with a low RI glass.   An improved photocathode of R12860 is expected to eanable it  to reach  the quantum efficiency of 30\% at 400 nm, about 1.4 times higher than that of the Super-Kamiokande  PMTs. The photoelectron collection efficiency of R12860 is also much higher. As a result,  the total efficiency for the single photon detection of Hamamatsu R12860 is almost twice higher than that of the Super-Kamiokande PMTs, as shown in Fig.~\ref{fig:hyperk_pmt}.
The charge resolution of single photoelectron of R12860 is evaluated to be 35\%,   while the transit time spread measured for single photoelectrons at a fixed threshold is about  2.6 ns (full width at half maximum).
To avoid a chain implosion of the  Hyper-Kamiokande PMTs in deep water a cover made  of  a stainless steel and a UV transparent acrylic  is developed. All tests in water at the depth of  up to 80 m   confirmed that the cover can be used to prevent chain implosions in Hyper-Kamiokande~\cite{Nishimura:2020eyq}.
Since the photon collection efficiency of a Box \& line PMT decreases by 
about 3\% at a magnetic field of about  200 mG  perpendicular to the PMT direction, a geomagnetic field compensation coil will be used to keep the remaining magnetic field perpendicular to  PMTs  smaller than 100 mG.  It is worth noting that  136 R12860 PMTs  were installed in Super-Kamiokande  instead of  failed PMTs during the  Super-Kamiokande  refurbishment in summer 2018. As tests showed, parameters of these R12860 PMTs met the requirements of the experiment~\cite{bronner}.
 
\begin{figure}[htbp]
\centering 
\includegraphics[width=10cm]{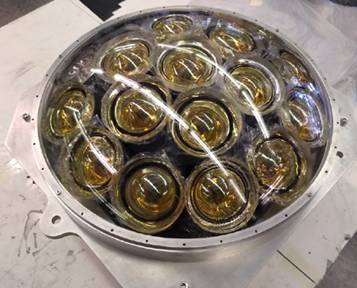} 
\caption{\label{fig:mpmt} Multi-PMT optical module consisting of nineteen  8 cm PMTs. }
\end{figure}

Hyper-Kamiokande performance can be improved by using new photosensors which are under development for other projects. For example, an interesting option could be a multi-PMT optical module (mPMT) based on the KM3NeT design~\cite{km3net}. The first prototype of such a module was built at TRUIMF~\cite{mpmt_triumf}. It comprises of 19  Hamamatsu R14374 8 cm PMTs, as shown in Fig.~\ref{fig:mpmt}.
Multi-PMTs are less sensitive to magnetic field and  provide better timing resolution (about 0.6 ns)  and  better vertex resolution near the detector walls.   

Large area PMTs based on  Micro Channel Plates (MCPs)~\cite{juno_mcp-pmt}  initially developed for the JUNO experiment  can be also used in Hyper-Kamiokande. MCP-PMTs are  produced by the  North Night Vision Technology Company in China. The original design of these photosensors was significantly improved: the transit time spread  was reduced from 14 ns to about 5 ns and  a charge peak resolution of about 43\% was obtained for a single photoelectron signal. The MCP-PMTs have  the   quantum efficiency of $\sim 30$\% at 400 nm that meets the requirements of Hyper-Kamiokande.   An important feature of these devices well suited for Hyper-Kamiokande is their low sensitivity to the magnetic field.
 
\section{Neutrino beam and the near detector complex}
\label{sec:near_detectors}

For oscillation measurements Hyper-Kamiokande will use  the same beam line as the  T2K experiment~\cite{Abe:2011ks} with the same off-axis configuration. The neutrino beam produced by  30 GeV protons extracted from the J-PARC accelerator  is aimed $2.5^{\circ}$ away from  Hyper-Kamiokande   as well as from Super-Kamiokande  to take advantage of the pion decay kinematics to produce a quasi-monoenergetic beam with a spectrum peaked at 600 MeV well adjusted to  the first oscillation maximum for a baseline of 295 km.  The  beam intensity of  more $2.6\times 10^{14}$ proton-per-pulse  (the pulse width is 5 $\mu s$, repetition period is 2.48 c)  has been achieved in T2K, corresponding to  the   515 kW beam power.   After the proton driver upgrade, the beam power of the main ring is expected to reach 1300 kW  with $3.2\times  10^{14}$ protons-per-pulse  and a 1.16 s repetition period  by 2027, i.e. before the start of  Hyper-Kamiokande.
 
The T2K  near   detector ND280~\cite{Abe:2011ks,Kudenko:2008ia}  measures neutrino beam  close enough to the  pion-production target so that oscillation effects are negligible. Measurements of  forward-going muons on the carbon target by the existing near detector are translated into constraints on the 4$\pi$ muon angular distribution on a water target seen at Super-Kamiokande. To reduce the systematic uncertainties to $\leq 4$\%  on the total event prediction in the far detector, in presence of oscillation, the ND280 near detector will be upgraded~\cite{Abe:2016tii} for T2K measurements before the start of Hyper-Kamiokande and then will be used for measurement of CP asymmetry in neutrino oscillations in Hyper-Kamiokande.  An intermediate water Cherenkov detector (IWCD) based on the design of the NuPRISM detector~\cite{nuprism} is proposed to be constructed at a distance of 1-2 km from the target. It will measure muon and electron neutrino (antineutrino) cross sections on water with the same solid angle as the far detector. A combination of data obtained  in  both  the magnetized ND280 detector and IWCD will help to further reduce systematic uncertainties in oscillation measurements.
 
\subsection{ND280 upgrade}
\begin{figure}[htbp]
\centering 
\includegraphics[width=.5\textwidth]{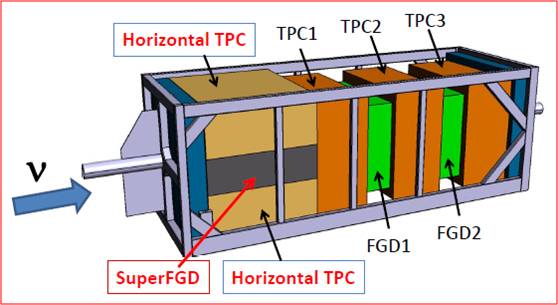}
\qquad
\includegraphics[width=.4\textwidth,origin=c,angle=0]{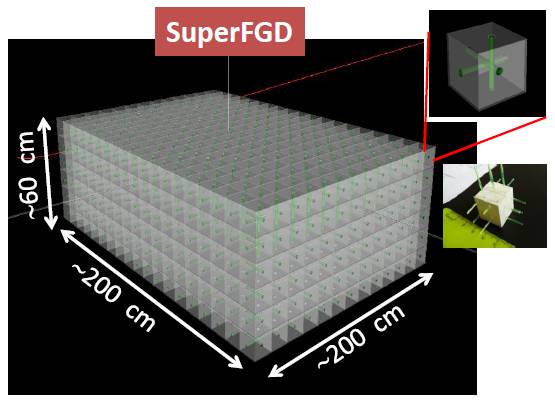}
\caption{\label{fig:nd280upgrade} 3D view of the ND280 detector upgrade (left) and   the SuperFGD structure (right). Also shown is a cube of  $1\times 1\times 1$ cm$^3$ with 3 orthogonal wave-length shifting fibers inserted into holes.}
\end{figure}

The upgrade keeps the current ND280 tracker, i.e. three vertical TPCs and two FGDs. The main part of the  P0D detector will be replaced by   a new highly granular fully active scintillator neutrino detector, two new TPCs, and  time-of-flight (TOF) planes.  The highly granular scintillator detector SuperFGD of a mass of about 2 tons  is comprised of $\sim 2\times 10^6$ small scintillator cubes with 1 cm side, each read out with WLS fibers in the three orthogonal directions coupled to compact photosensors,  Micro Pixel Photon Counters. SuperFGD will serve  as an active neutrino target and a $4\pi$ detector of charged particles from neutrino interactions.  SuperFGD is sandwiched between two High-Angle TPCs, readout by resistive Micromegas detectors, with a compact and light field cage. These detectors are surrounded by six large TOF planes to determine the track direction and improve the particle identification, as shown in Fig.~\ref{fig:nd280upgrade}.
SuperFGD (see Fig.~\ref{fig:nd280upgrade} (right)) is an innovative device with excellent detector performance. The beam tests at CERN PS showed that  a MIP crossing a single cube  produces about 40  photoelectrons per WLS fiber in realistic conditions. The timing resolution per fiber is obtained to be better than 1 ns~\cite{Mineev:2019dpe,Mineev:2018ekk}.  SuperFGD has a very good capability to track muons, pions and protons stopping in this detector over 4$\pi$ solid angle. Moreover its high granularity will allow us to distinguish electrons produced by electron neutrino interactions from converted photons. Studies are ongoing to evaluate the SuperFGD potential to detect neutrons.  Beam tests of a High-Angle TPC prototype at CERN  also showed good performance of resistive Micromegas detectors: excellent uniformity of the gain, a deposited energy resolution $dE/dx$  of about 9\%, and a spatial resolution of better than 300 $\mu m$ were obtained~\cite{Attie:2019hua}. 

\subsection{Intermediate water Cherenkov detector}
\begin{figure}[htbp]
\centering 
\includegraphics[width=11cm]{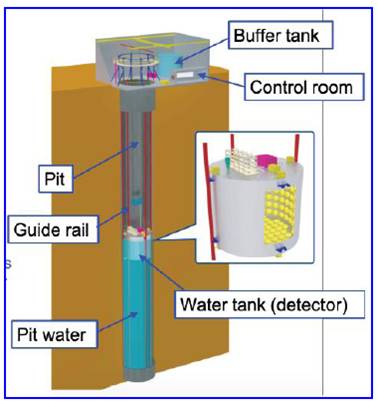} 
\caption{\label{fig:iwcd} The configuration of the intermediate water Cherenkov detector. The instrumented section of the tank moves vertically to cover different  off-axis angle regions.}
\end{figure}

The baseline design of IWCD considers the  location of the detector  at about  1 km downstream of the neutrino interaction target in a 50 m deep shaft. The detector must span the off-axis range $1^{\circ} - 4^{\circ}$ and  its diameter should be   large enough to contain the  required muon momentum of 1 GeV/c. This corresponds to a 50 m tall tank with a 6 m diameter inner detector  and a 10 m diameter outer detector, as shown in Fig.~\ref{fig:iwcd}.
The novel feature of this detector is the ability to raise and lower the instrumented section of the tank in order to span the full off-axis range. The inner detector will be instrumented with multi-PMT optical modules described in Section~\ref{sec:pmts}.  Compact size and high timing resolution of 8 cm PMT's will allow us to improve the vertex resolution and  particle identification in comparison with Box{\&}Line PMTs. 

\section{Conclusion}
\label{sec:conclusion}

The Hyper-Kamiokande project is officially approved  and the detector is   expected to be constructed and ready for physics measurements in 2027.  This new experiment  is based on the experience and facilities of the already existing and very successful Super-Kamiokande and  T2K and will use novel photosensors for detection of the Cherenkov light.    The J-PARC proton accelerator will be  upgraded to reach a MW beam for Hyper-Kamiokande. A complex  of near detectors which includes the  upgraded ND280 detector and  IWCD  will be of great importance for a sensitive search for CP violation in neutrino oscillations.  Hyper-Kamiokande  will be a multipurpose neutrino detector with a rich physics program that includes the observation of the leptonic  CP violation,   a  search for the proton decay, detection neutrinos  from Supernova, and  astrophysical neutrinos.

\acknowledgments

This work was supported  by the RFBR-JSPS  grant \# 20-52-50010.


\end{document}